\documentclass{article}
\usepackage[bibtex,parskip]{general}

\newacronym{mimo}{MIMO}{multiple input multiple output}
\newacronym{bibo}{BIBO}{bounded input bounded output}

\begin{document}
\title{Network Systems and String Stability}
\author{S. St\"{u}dli, M. M. Seron, and R. H. Middleton}
\maketitle

 \begin{abstract}
 Network systems and their control are highly important and appear in a variety of applications, including vehicle platooning and formation control. Especially vehicle platoons are highly investigated and an interesting problem that arises in this area is string stability, which broadly spoken means that a input signal amplifies unbounded as it travels through the vehicle string. However, various definitions are commonly used. In this paper, we aim to formalise the notion of string stability and illustrate the importance of those distinctions on simulation examples.  A second goal is to generalise the found definitions for general network systems.
 \end{abstract}

\section{Introduction}

Networked systems and their control are studied in a variety of fields, such as 
vehicular platooning \cip{levine1966vehiclePlatoon,Melzer1971vehiclePlatoon,peppard1974pid,rick2010stringinstability,seiler2004vehicleStrings,barooah2005string,swaroop1996stringstability,swaroop1999ahs,martinec2016assymetric,herman2015fielder,lestas2007vehiclePlatoon,cook2007vehiclePlatoon,rogge2008vehicleplatoons}, 
formation control \cip{fax2004platoons,zelazo2008formation,yadlapalli2006stability}, 
and many others. These systems all consist of many agents that are performing a common task. While in early stages centralised controllers were studied \cip{levine1966vehiclePlatoon,Melzer1971vehiclePlatoon}, such controllers become infeasible if the number of agents increases. Hence, distributed and decentralised approaches are investigated, where the agents utilise local information and in some cases information transmitted by other agents \cip{peppard1974pid,rick2010stringinstability,seiler2004vehicleStrings,barooah2005string,fax2004platoons,li2011performance,tonetti2010performancelimits,tonetti2011performance,swaroop1999ahs,martinec2016assymetric,herman2015fielder,cook2007vehiclePlatoon,yadlapalli2006stability}. 

However, in some cases these distributed systems experience undesired properties such as instability, amplification of disturbances within the network, and cascading failures. It is therefore of utmost importance to understand the dynamics and limitations that are imposed on these systems with respect to the information flow as well as the underlying systems. In this work we will mostly concentrate on instability and amplification of signals, such as input disturbances and review some of the main results in this field.

\begin{remark}
  A related field of research is that of consensus algorithms, where multiple agents aim to equalise a state variable \cip{moreau2005consensus,saber07consensus,zelazo2007edgeagreement}. Normally, the dynamics of this state variable is simple, for example its derivative is set directly to a weighted average of the state variables of the neighboring agents. Nonetheless, the similarities between the two allow the use of similar techniques, such as graph theory. 
\end{remark}

The methods used to analyse these systems range from classical control theory to spatial-temporal systems \cip{bamieh2002distributedcontrol,steffi2012thesis,steffi2014passivity}. 
Recent works combine control theoretic approaches with graph theory. In that context the agent's behaviour is governed by an individual dynamic system, while the information exchange among the agents is represented as a graph. The behaviour of the system is then closely linked to the Laplacian of the graph and in more detail its eigenvalues, such that the study of the Laplacian becomes an integral part in the analysis of the networked system \cip{tonetti2010performancelimits,tonetti2011performance,li2011performance,herman2015fielder,you:2013:ncs_survey,barooah2005string,yadlapalli2006stability}. While some works do not consider this link to graph theory directly, their problem description and to some extent the results can be translated into the same separation of the agents individual dynamics and the graph considering the information exchange, see for example \cit{seiler2004vehicleStrings,rick2010stringinstability}. 

In this context there are three important design choices to consider besides the actual controller design:
\begin{enumerate*}[itemjoin={{; }},itemjoin={{; and}},after=.]
\item the dynamic system representation, linear vs. non-linear
\item the individual agent's dynamic system, heterogeneous vs. homogeneous
\item the communication structure among the agents
\end{enumerate*}

The individual dynamic system of each agent is often considered to be linear. In that case it is represented either as a transfer function \cip{tonetti2011performance,tonetti2010performancelimits,rick2010stringinstability,martinec2016assymetric,cook2007vehiclePlatoon,barooah2005string,yadlapalli2006stability} or in state space \cip{li2011performance,herman2015platoons,herman2015circular,rogge2008vehicleplatoons}. Both forms of representing the dynamic system allow the usage of specific system analysis and control techniques and both are common in the literature, where in some instances  both representations are used in combination to extend or facilitate the results \cip{fax2004platoons,herman2015fielder,cook2005stability}. As seen in \cit{cook2005stability} the two approaches are in effect interchangeable for the analysis of string instability in vehicle platoons. In the linear case we hence investigate individual system dynamics of similar forms to
\begin{equation}
  Y_{i}(s) = \underbrace{P_{i}(s) K_{i}(s)}_{L_{i}(s)} U_{i}(s) 
\end{equation}
with an individual controller $K_{i}(s)$ and system dynamics $P_{i}(s)$ for vehicles $i \in \set{1,\ldots,N}$ and assume $0$ initial conditions. As commonly known if $K_{i}(s)P_{i}(s)$ is a  proper transfer function a state space representation can be found, however it is not unique. In this context we look at a linear representation of the form 
\begin{subequations}
  \begin{gather}
    \dot{x}_{i}(t) = A_{i} x_{i}(t) + B_{i} u_{i}(t) \\
    y_{i}(t) = C_{i} x_{i}(t) + D_{i} u_{i}(t),
  \end{gather}
\end{subequations}
where $x_{i}(t)$ is the state of the \nth{i} subsystem and $u_{i}(t),y_{i}(t)$ are the time signals corresponding to the Laplace transforms $U_{i}(s)$ and $Y_{i}(s)$. The signals $y_{i}(t)$ are the outputs of the system or some chosen performance variables, for example in a platoon this is often the inter-vehicle spacing error. The signals $u_{i}(t)$ are the inputs to the controller and includes the information from other agents as well as a given reference. Hence the subsystems will be linked through these inputs. 

Sometimes the use of non-linear models and controllers is preferred. The reason thereof is partly the fact that in many networked systems the dynamics are inherently non-linear and linear models make use of linearisation around an operating point. While for a broad range of applications the linearisation works well within the regulated bounds, once the operation deviates considerable from the operating point the dynamics no longer represent the system accurately enough. Secondly, it is possible to avoid certain undesirable effects by the use of non-linear control approaches \cip{yanakiev1998nonlinear}. While, we mainly focus on linear dynamics we will include some comments regarding non-linear approaches where appropriate. 

Independent of the nature of the dynamic system, it is important to distinguish two approaches for the dynamic systems of all considered agents: 
\begin{enumerate}
\item homogeneous agents, \acs{ie} the dynamic systems and their controllers of all agents are identical (this means in the linear case $P(s)$ and their controllers $K(s)$ are identical) \cip{tonetti2010performancelimits,martinec2016assymetric,herman2015fielder,herman2015platoons,herman2015circular,cook2007vehiclePlatoon,barooah2005string,seiler2004vehicleStrings,yadlapalli2006stability};
\item heterogeneous agents, \acs{ie} the dynamic systems and/or their controllers vary among the agents (this means in the linear case $P_{i}(s)$ or their controllers $K_{i}(s)$ are not necessarily equal) \cip{tonetti2011performance,rick2010stringinstability,dunbar2006formation,lestas2007vehiclePlatoon}.
\end{enumerate}
The use of homogeneous agents simplifies the analysis, however idealises the systems dramatically. Hence, it is important to extend the results where possible to heterogeneous networked systems. This is also important in regard to model uncertainties and small model changes that are undoubtedly present. In some cases the use of heterogeneous controllers is even suggested to improve the performance of the system \cip{khatir2004platoon}.

The other main design choice relates to the interaction among the agents and hence the graph structure that is used. These graphs are either un-directed \cip{li2011performance,yadlapalli2006stability} leading to symmetric, bi-directional communication structures, or directed \cip{tonetti2010performancelimits,tonetti2011performance,martinec2016assymetric,herman2015fielder,rogge2008vehicleplatoons} allowing for asymmetric control structures that can improve the performance of the system. Further, some works include weighted edges \cip{herman2015fielder,herman2015circular}, which allows for more complex control strategies. The graph is commonly described using its Laplacian, which we denote $L$ in the following.

Using the Laplacian $L$, we can write the control input as
\begin{equation}
  u(t) = L (y_{ref}(t) - y(t)),
\end{equation}
where $u(t)$ and $y(t)$ are the vectors containing the inputs and measured outputs of each agent and $y_{ref}(t)$ is a reference signal. For example, in vehicle platooning $y(t)$ could be the inter-vehicle spacing or the position of the vehicles. $Y_{ref}(s)$, $Y(s)$, and $U(s)$ denote their Laplace transforms. Then, the networked system can be described by
\begin{multline}
  Y(s) = \invert{\left(\Id + \diag(P_{i}(s))\diag(K_{i}(s)) L \right)}\\ \diag(P_{i}(s))\diag(K_{i}(s)) L Y_{ref}(s).
\end{multline}
We denote the transfer matrix from the reference input to the measured outputs of the networked system $H(s)$. To specify certain entries of this matrix we use subscripts, such as $H_{i,j}(s)$, where we use $:$ to indicate all entries, \acs{ie} if we refer to a row or column of $H(s)$.

\begin{remark}
In some cases the reference signal and the measured outputs are not chosen to represent the same variable, which means that the Laplacian will be split in two parts where the first part describes the mapping between the measured output and reference and the second part describes the control configuration. For example in the case of vehicle platoons, the reference could be the inter-vehicle position, while the measured output is the position of the vehicles \cip{rick2010stringinstability}. The advantage of such a formulation can be a simpler formulation of the reference input to the system.
\end{remark}

Equivalently, the networked system can be expressed in state space form as 
\begin{subequations} 
  \label{eq:system-equation-ss-general}
  \begin{gather}
    \begin{multlined}
      \dot x(t) =\left( \diag(A_{i}) - \diag(B_{i}) L \diag(C_{i})\right) x(t) \\ + \diag(B_{i}) L y_{ref}(t)
    \end{multlined}\\
   y(t) = \diag(C_{i}) x(t). 
  \end{gather}
\end{subequations}
Therefore, the Laplacian $L$ is very important for the analysis of such networked systems and many works investigate the eigenvalues of $L$ and their influence on the whole system \cip{herman2015circular,you:2013:ncs_survey,barooah2005string}. 

\begin{remark}
  It is common to investigate the response to a disturbance rather than a reference signal change. In that case we investigate the transfer matrix $G(s)$ as found in 
\begin{multline}
  Y(s) = \underbrace{\invert{\left(\Id + \diag(P_{i}(s))\diag(K_{i}(s)) L \right)}}_{G(s)} D(s),
\end{multline}
where $D(s)$ is the Laplace transform of disturbances acting on the vehicles or in case of state space models
\begin{subequations} 
  \label{eq:system-equation-ss-general}
  \begin{gather}
    \begin{multlined}
      \dot x(t) =\left( \diag(A_{i}) - \diag(B_{i}) L \diag(C_{i})\right) x(t) \\ + \diag(B_{i}) L d(t)
    \end{multlined}\\
   y(t) = \diag(C_{i}) x(t) + d(t). 
  \end{gather}
\end{subequations}

\end{remark}

This approach can be extended by allowing multiple Laplacians for different inputs to the vehicle \cip{herman2015circular} or by allowing $P_{i}(s)$ and $K_{i}(s)$ to be transfer function matrices \cip{fax2004platoons}. Such configurations become naturally more complex, however give more freedom to the controller and can improve the performance of the networked system. 

Another possible extension is to consider time dependent information exchanges which would result in time dependent Laplacians. While this is investigated for consensus systems \cip{moreau2005consensus}, to the best of our knowledge it has not been investigated directly for networked systems as considered here.

An even newer approach for the analysis of networked systems utilises the so called wave approach \cip{martinec2016assymetric,herman2015platoons,herman2015circular}. The idea is to model the state of the system as waves propagating through the graph structure. While this method is mainly used in relation to vehicle platooning, the approach can be extended for other structures \cip{martinec2016assymetric}.

While a lot of work is focused on the effects of using an idealised communication set-up, there are some works that investigate properties when communication constraints occur, such as delays or losses, \cip{oncu212stringstability,andres2015thesis,andres2014delays,seiler2001losses}. These additional constraints make the control more difficult and can degrade the performance of the controller considerably. These issues become more prominent when the communication occurs over a shared network, where there are additionally bandwidth limitations and packet drop outs. These issues worsen the conditions and impair both state estimation and control of such systems. There is the field of networked control, see \cit{hespanha:2007:survey,you:2013:ncs_survey,yang2006ncs} and references therein for a summary, that deals solely with these problems. In this work, we concentrate  instead on limitations of the performance caused by the communication structure and the individual systems. Both approaches are of importance for the future utilisation of networked systems.

In this paper, we will focus mostly on vehicle platooning, which is a special case of networked systems that have widely been studied. In this regard, a property denoted string stability is commonly mentioned as a performance criteria for vehicle platoons. We will in \cref{sec:vehicle-platoons} formalise different variations commonly used of this property that while related contain subtle differences. Additionally, we will discuss the results found for vehicle platoons in regard to stability as well as the property of string instability. In \cref{sec:gener-netw-syst} results for more general systems are revised that illustrate how the property of string stability can be extended to such more general systems. Finally, in \cref{sec:conclusion} we will conclude our findings and pinpoint to some open problems in this field.

\section{Vehicle Platoons}
\label{sec:vehicle-platoons}

One of the applications that is very well studied is vehicle platooning sometimes referred to as automated highway systems. In this application multiple vehicles driving in a string are controlled to keep a defined distance to each other. This defined distance is often mentioned as spacing policy and is in some cases used as a design parameter of the system. The two most commonly used policies are either constant distance, \acs{ie} an absolute distance between the vehicles, and constant time headway, a velocity dependent distance between the vehicles. In \cref{sec:struct-prop}, we introduce the usual structural properties that are present in this application. Then, we introduce in \cref{sec:stability-platoons,sec:string-stability-platoons} the properties of stability and string stability in this context. Finally, in \cref{sec:results} we revisit the main contributions in this field. 

\subsection{Structural properties}
\label{sec:struct-prop}

The individual open loop dynamics of such systems, consisting of the vehicle model and the controller, contain usually two integrators, which allows them to follow given ramp inputs. Hence, in the linear case we investigate open loop dynamics of single agents of the form
\begin{equation}
  L_{i}(s) = \frac{N_{i}(s)}{s^{2}D_{i}(s)}
\end{equation}
if considering transfer functions. Note that the double integrator also introduces some limitations on the state space representation $A_{i}$. 

Further, these systems lead to a special structure of the graph that represents the information exchange. There are two main categories; the first is a string connection where the first vehicle is independent or following a virtual leader \cip{rick2010stringinstability,martinec2016assymetric,seiler2004vehicleStrings,andres2015thesis,cook2007vehiclePlatoon}, the second is a cyclic interconnection where the first vehicle uses the last vehicle as its predecessor \cip{herman2015circular,andres2015thesis,rogge2008vehicleplatoons}. 

The first category leads to graphs that form chains, such that the associated adjacency matrix is a band diagonal matrix. In some cases additional communication links are present that disturb this band diagonal matrix structure. For example \cit{seiler2004vehicleStrings} broadcasts the leader position additionally to local information to avoid disturbance amplification. 

The second category leads to cyclic graphs where the adjacency matrix is a circulant matrix. Systems that can be represented in such a form are for example light rail or subway circuits. Further, such circular structures can be useful to establish results for the non-cyclic case, since the analysis is simplified. Such an approach is followed in \cit{herman2015platoons,herman2015circular}.

\subsection{Stability}
\label{sec:stability-platoons}

As is common in control theory the closed loop stability of the vehicle string is important. Here, unlike in normal conditions the stability is often investigated not just for a given string length but for increasing number of vehicles, \acs{ie} a system can become unstable for large enough $N$ \cip{barooah2005string,rogge2008vehicleplatoons,andres2015thesis}. We will use the term eventual instability to denote situations where instability occurs for long enough vehicle strings. Otherwise stability is understood in the usual sense of asymptotic stability or \ac{bibo} stability in the case of input-output models. 

\subsection{String stability}
\label{sec:string-stability-platoons}

\begin{figure*}[htb]
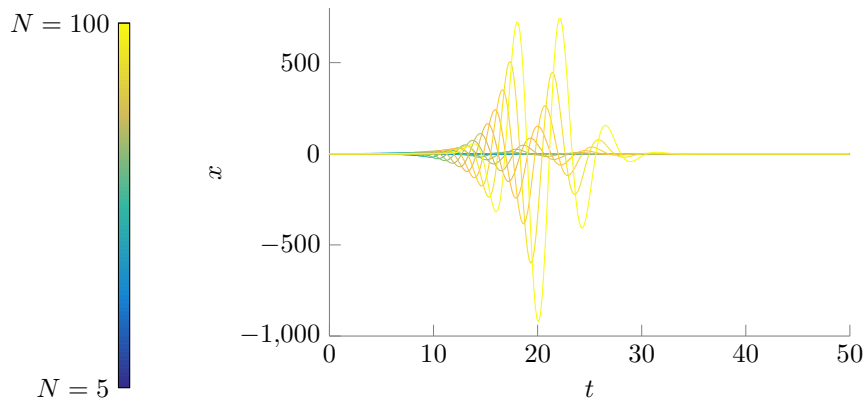

  \centering  
  \addcolourbar{$N=5$,$N=100$}
  \includepgf[][0.6\textwidth]{example1-stringinstability}
  \caption{Example for a string unstable system that uses homogeneous agents with open loop transfer function $L(s)=\frac{6 s + 5}{s^2}$ and a uni-directional communication structure. The plot shows the response of the last vehicle in the string to an impulse disturbance acting on the first vehicle. 
  }
  \label{fig:example1-stringinstability}
\end{figure*}

Once the stability of the vehicle string is assured, \acs{ie} it is not eventually unstable, string stability is an important characteristic of vehicle strings, which in a broad sense means that a disturbance will not amplify when propagating through the string \cip{peppard1974pid}. Besides disturbances the effect of reference changes or other inputs to the system can be considered. String stability is a very different effect from eventual stability. While eventual instability means that the interconnected system for some length $N$ looses stability, this does not occur for a string unstable system. However, the magnitude of the response, \acs{ie} the amplification of an input signal, will increase as the string length increases. \Cref{fig:example1-stringinstability} shows a typical response to an impulse disturbance acting on the first vehicle for a string unstable system that is not eventual unstable. The response with the smallest peaks (shown in dark green) is that of the last vehicle when the string consists of $5$ vehicles whilst the response with the largest peaks (in light yellow) corresponds to the response of the last vehicle when the string consists of $100$ vehicles. 

There are different definitions of string stability, for example based on the $H_{\infty}$ norm \cip{rick2010stringinstability,peppard1974pid,eyre1998stringStability,hermannThesis} or the $L_{\infty}$ signal norm \cip{swaroop1996stringstability,swaroop1999ahs,canudas1999separation,rogge2008vehicleplatoons}, and related stability issues such as flock stability \cip{cantos2014transients,herman2015platoons}, harmonic stability \cip{herman2015fielder}, and eventual string stability \cip{khatir2004platoon} in the literature. \cit{swaroop1996stringstability} gives a more formal definition of string instability based on the $L_{\infty}$ signal norm with respect to initial conditions. \cit{rogge2008vehicleplatoons} extends these definitions, which lead to a more strict form of string stability that includes avoidance of the so called slinky effect, \acs{ie} amplification from one vehicle to the next. Besides this extension they give a more formal definition of string stability with respect to input signals. \cit{cook2005stability} gives formal definitions of, so called practical string stability, in terms of more general signal norms. Note that depending on the actual definition of string stability some systems achieve one form of string stability while they remain ``unstable'' in another sense. This fact makes the comparison of different results more challenging due to the various nuances present. 

We now formalise the definitions of the most commonly used forms of string stability, including the ones mentioned above.
Note that all the definitions have in common that the amplification of a signal does not approach infinity with increasing string lengths. The main distinction among the various versions occurs in which signals are considered and the characteristics of the bound on the amplification. While these differences are small they are very important to gain an overall accurate analysis of the system. In the following we will look at these distinctions and their importance. To this end we first introduce formally various versions of input-output string stability, where the main difference lies in the considered inputs and outputs.  Afterwards, we will introduce the very important notion of internal stability that is commonly used. Finally, we will look at other notions of string stability that allow a more precise classification. 


\begin{table*}[htb]
  \centering
  \begin{tabularx}{\linewidth}{p{9em}Xl}
\thead{ & Definition & used in} 
 
  general & If for every input signal $z(t)$ with $\norm[p]{z}<\infty$ there exists $R$ for all $N \in \N$  such that \begin{equation*} \norm[q]{y} < R. \end{equation*}  & \cip{barooah2009mistuning,seiler2004vehicleStrings}\\

  single with input $\alpha$ & If for every input signal $z(t)$ with $z_{i}(t) = 0$ for $i \neq \alpha$ and $\norm[p]{z_{\alpha}}< \infty$ there exists $R$ for all $N \in \N$ such that \begin{equation*} \norm[q]{y} < R. \end{equation*}  & \cip{barooah2005string} \\

  single & If for every $i$ and input signal $z(t)$ with $z_{j}(t) = 0$ for $j \neq i$ and $\norm[p]{z_{i}}<\infty$ there exists $R$ for all $N \in \N$  such that \begin{equation*} \norm[q]{y} < R. \end{equation*}  & \\

  final & If for every input signal $z(t)$ with $\norm[p]{z}<\infty$ there exists $R$ for all $N \in \N$  such that \begin{equation*} \norm[q]{y_{N}} < R. \end{equation*}  & \cip{andres2015thesis}\\

  single with input $\alpha$ final & If for every input signal  $z(t)$ with $z_{i}(t) = 0$ for $i \neq \alpha$ and $\norm[p]{z_{\alpha}}<\infty$ there exists $R$ for all $N \in \N$  such that \begin{equation*} \norm[q]{y_{N}} < R. \end{equation*}  &  \cip{rick2010stringinstability,khatir2004platoon}\\
  
  single final & If for every $i$ and input signal $z(t)$ with $z_{j}(t)=0$ for $j \neq i$ and $\norm[p]{z_{i}}<\infty$ there exists $R$ for all $N \in \N$  such that \begin{equation*} \norm[q]{y_{N}} < R. \end{equation*}  &  \cip{rick2010stringinstability,khatir2004platoon}\\

\bottomrule
  \end{tabularx}
  \caption{$l_{p,q}$ string stability or $l_{p}$ string stability if $p=q$. Most commonly $p=q=2$ or $p=q=\infty$ is used. Various definitions are collected differentiated by the input, outputs, and states considered. The input signals $z$ are most commonly either disturbances or reference changes, while the output signals $y$ are most commonly the inter-vehicle spacing errors between the vehicles. Note that the signals $z_{i}$ or $y_{i}$ denote the inputs and outputs acting on vehicle $i$, respectively.}
  \label{tab:string-stability}
\end{table*}

\begin{table*}[htb]
  \centering
  \begin{tabularx}{\linewidth}{p{9em}Xl}
    \toprule
    Stability notion & Definition & used in \\ \midrule

    harmonic  & For every input signal $z(t)$ with $\norm[p]{z} < \infty$ there exists $R$ such that \begin{equation} \norm[q]{y} < R^{N} \mathskip \forall N. \end{equation} & \cip{herman2015fielder,cantos2014transients}\\

strict & Assume a vehicle platoon with a clear defined single leader and followers. Then, the platoon is strict $l_{p,q}$ string stable from $\bar{N}$ if it is $l_{p,q}$ string stable and for any input signal $z_{i}(t)$ with $\norm[p]{z_{i}} < \infty$ and vehicles $j$ and $k$, where $j$ is any vehicle following $i$ and $k$ is a direct follower of $j$,  
\begin{equation}
  \norm[q]{y_{k}} \leq \norm[q]{y_{j}} 
\end{equation}
for all $j\geq \bar{N}, k$.  & \cip{peppard1974pid,rogge2008vehicleplatoons} \\

weak & For any $\epsilon>0$ there exists $\delta(\epsilon)>0$ (independent of $N$) such that for any input $\norm[p]{z}<\delta(\epsilon)$ it follows that
\begin{equation*}
  \norm[q]{y} < \epsilon \mathskip \forall N.
\end{equation*}
& \cip{rogge2008vehicleplatoons,steffi2014passivity,steffi2009timeheadway,swaroop1996stringstability,canudas1999separation,cook2005stability,steffi2008diplomarbeit}\\
    \bottomrule
  \end{tabularx}
\caption{Different notions of string stability.}
\label{tab:other-string-stability-forms}
\end{table*}


\begin{table*}[htb]
  \centering
  \begin{tabularx}{\linewidth}{p{14em}X}
    \toprule
    Stability & Conditions  \\ \midrule

    $l_{2}$ string stability & The $H_{\infty}$ norm of a \acs{mimo} transfer matrix $X(\jm \omega)$, normally $H(s)$ or $G(s)$, is bounded independently of the string length, \acs{ie}    \begin{equation}  \sup_{\omega}(\sigma_{max}{(X(\jm \omega))} < \infty \end{equation}  for all $N \in \N$\\ 

single $l_{2}$ string stability & The $H_{\infty}$ norm of a transfer matrix $X_{:,1}(\jm \omega)$, normally $H_{:,1}(s)$ or $G_{:,1}(s)$, is bounded independently of the string length, \acs{ie}
$\norm[\infty]{X_{:,1}} < \infty$ for all $N \in \N$ \\

    single final $l_{2}$ string stability & The $H_{\infty}$ norm of a transfer function $X(\jm \omega)$ from a single input to the output of the last vehicle, normally $H_{1,n}(s)$ or $G_{1,n}(s)$, is bounded independently of the string length, \acs{ie} \begin{equation*}  \sup_{\omega}(\betrag{X(\jm \omega)}) < \infty. \end{equation*}  for all $N \in \N$ \\

    \bottomrule
  \end{tabularx}
  \caption{Conditions for string stability}
  \label{tab:string-stability-cond}
\end{table*}


\begin{table*}[htb]
  \centering
  \begin{tabularx}{\linewidth}{p{9em}Xl}
\thead{ & Definition & used in} 

  general & If for every input signal $z(t)$ with $\norm[p]{z}<\infty$ there exists $R$ for all $N \in \N$  such that \begin{equation*} \norm[q]{y} < R. \end{equation*}  \\

  single input & If for every $i \in \set{1,\dots,N}$ and input signal $z(t)$ with $z_{j}(t) = 0$ for $j \neq i$ and $\norm[p]{z_{i}}<\infty$ there exists $R$ for all $N \in \N$  such that \begin{equation*} \norm[q]{y} < R. \end{equation*}  & \\

  single output  & If for every input signal $z(t)$ with $\norm[p]{z}<\infty$ there exists $R$ for all $N \in \N$  such that \begin{equation*} \norm[q]{y_{\ell}} < R \mathskip \forall \ell \in \set{1,\dots,N} \end{equation*}  & \\

  single input-output  & If for every $i \in \set{1,\dots,N}$ and input signal $z(t)$ with $z_{j}(t) = 0$ for $j \neq i$ and $\norm[p]{z_{i}}<\infty$ there exists $R$ for all $N \in \N$  such that \begin{equation*} \norm[q]{y_{\ell}} < R \mathskip \forall \ell \in \set{1,\dots,N} \end{equation*}  & \\

\bottomrule
  \end{tabularx}
\caption{$l_{p,q}$ networked stability or $l_{p}$ networked stability if $p=q$. Various definitions are collected differentiated by the input, outputs, and states considered.  }
  \label{tab:networked-stability}
\end{table*}

The notion of $l_{p,q}$ string stability is presented in \cref{tab:string-stability}. While we concentrate here on $l_{p,q}$ signal norms, other norms could possibly be considered such as power signal norms. However, to the best of our knowledge the $l_{2}$ and $l_{\infty}$ norms are by far the most considered ones. Especially, the relation between the $l_{2}$ signal norms and the $H_{\infty}$ system norm are commonly used for the investigation of string stability in linear systems. To this extent we collected some conditions for $l_{2}$ string stability variations in \cref{tab:string-stability-cond} in terms of transfer functions that were used  previously in similar form to define string stability. The reasoning behind these conditions follow directly from the definitions of the $l_{2}$ signal norm and the $H_{\infty}$ system norm. Note that similar conditions can be obtained for $l_{2,\infty}$ string stability in terms of the $H_{2}$ system norm. 

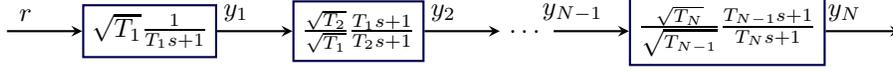
\begin{figure*}[htb]
  \centering
  \begin{tikzpicture}
    \coordinate (start);
    \node[bd,right = of start](a){$\sqrt{T_{1}} \frac{1}{T_{1}s+1}$};
    \node[bd,right = of a](b){$\frac{\sqrt{T_{2}}}{\sqrt{T_{1}}} \frac{T_{1}s+1}{T_{2}s+1}$};
    \node[right = of b](c){$\dots$};
    \node[bd,right = of c](n){$\frac{\sqrt{T_{N}}}{\sqrt{T_{N-1}}} \frac{T_{N-1}s+1}{T_{N}s+1}$};
    \coordinate[right = of n] (end);

    \draw[bdLine] (start) --node[above,near start]{$r$} (a);
    \draw[bdLine] (a) --node[above,near start]{$y_{1}$} (b);
    \draw[bdLine] (b) --node[above,near start]{$y_{2}$} (c);
    \draw[bdLine] (c) --node[above,near start]{$y_{N-1}$} (n);
    \draw[bdLine] (n) --node[above,near start]{$y_{N}$} (end);
  \end{tikzpicture}
  \caption{Example for $l_2$ vs $l_\infty$ string stability, see also Fig. 1.1 in \cit{steffi2008diplomarbeit}}
  \label{fig:example-l_2-vs}
\end{figure*}

\begin{figure*}[htb]
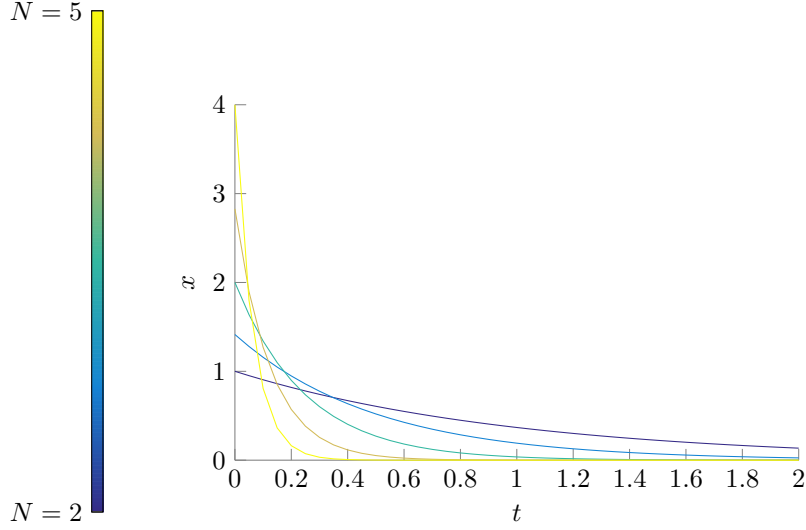

  \centering
  \addcolourbar[0.55\textwidth]{$N=2$,$N=5$}
  \includepgf[][0.65\textwidth]{example3-infty2stringstability}
  \caption{Example for $l_2$ vs $l_\infty$ string stability, see also Fig. 1.2 in \cit{steffi2008diplomarbeit}.}
  \label{fig:example-l_2-vs-1}
\end{figure*}

We will now first illustrate the importance of the signal norm considered. Therefore, we reproduce an example given in \cit{steffi2008diplomarbeit}, which shows that $l_{2}$ and $l_{\infty}$ string stability are not equivalent. To this end a non-homogeneous string of vehicles is interconnected as in  \cref{fig:example-l_2-vs}. The parameters are chosen such that $T_{i} > T_{i+1}$ holds for all $i$, which is achieved by using $T_{i} = \left(\frac{1}{2}\right)^{i-1}$. Setting the initial states $0$ except for the first vehicle which is set to $1$ produces the initial condition response plotted in \cref{fig:example-l_2-vs-1}, where the dark blue shows the result for a string with length $2$ and the bright yellow for a string with length $5$. This shows that while the system is $l_{\infty}$ string unstable, since the initial peak increases with increasing number of vehicles in the string, it retains $l_{2}$ string stability, since the decay speed increases fast enough with increasing number of vehicles in the string. We refer the reader to \cit{steffi2008diplomarbeit} for a detailed mathematical analysis of the example.

\begin{figure*}[ht]
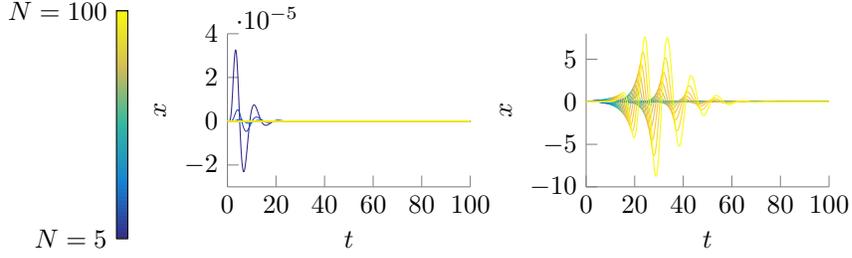

  \centering
  \addcolourbar[0.25\textwidth]{$N=5$,$N=100$}
  \begin{subfigure}[t]{0.35\textwidth}
    \includepgf{example2-singlefinalstring-stability}
    \caption{Last vehicle response to a disturbance acting on the first vehicle. }\label{fig:example2-singlefinalstring-stability}
  \end{subfigure}\quad
  \begin{subfigure}[t]{0.35\textwidth}
    \includepgf{example2-singlefinalstring-instability}
    \caption{First vehicle response to a disturbance acting on the last vehicle. }\label{fig:example2-singlefinalstring-instability}
  \end{subfigure}
  \caption{Example for a system that experiences single final string stability, but not general string stability. The system uses homogeneous agents with open loop transferfunction $L(s) = \frac{6s+5}{s^2}$ and a bi-directional communication structure.}
  \label{fig:example2-singlefinalstring}
\end{figure*}

The main distinction between the different variations of $l_{p,q}$ string stability are the considered inputs and outputs, see \cref{tab:string-stability}. To illustrate that it is important to consider these different variations, we show an example of a system that exhibits single with input $1$ final string stability, however is clearly general string unstable. \Cref{fig:example2-singlefinalstring} shows the responses of a vehicle platoon with increasing $N$ to a disturbance input. First, we assume that the disturbance acts on the first vehicle alone and we are interested in the last vehicle's response, which is shown in \cref{fig:example2-singlefinalstring-stability}. From this,  we see that the system is single with input $1$ final string stable, since with increasing $N$ the peak does not grow. However, this observation alone does not guarantee that the system is general string stable or even single string stable. This becomes clear when we assume that the disturbance acts on the last vehicle instead and we are interested in the effect on the first vehicle, which is shown in \cref{fig:example2-singlefinalstring-instability}. Here, the peak in the position clearly increases with increasing $N$ which means the system is general string unstable as well as single string unstable. 

\begin{figure}[ht]
  \centering
  \begin{tikzpicture}
    \node[draw, ellipse] (general){general};
    \node[draw, ellipse,below right = of general] (final){final};
    \node[draw, ellipse,below left = of general] (single){single};
    \node[draw, ellipse,below = 3cm of general] (singlefinal){single final};

    \draw[double,->] (general) --node[left]{1} (final);
    \draw[double,->] (general) --node[left]{2} (single);
    \draw[double,->] (general) --node[left]{3} (singlefinal);
    \draw[double,->] (single) --node[left]{4} (singlefinal);
    \draw[double,->] (final) --node[left]{5} (singlefinal);
  \end{tikzpicture}
  \caption{Overview over relations among different variations of $l_{p}$ string stability.}
  \label{fig:networkedSystems:cent-capt-over}
\end{figure}
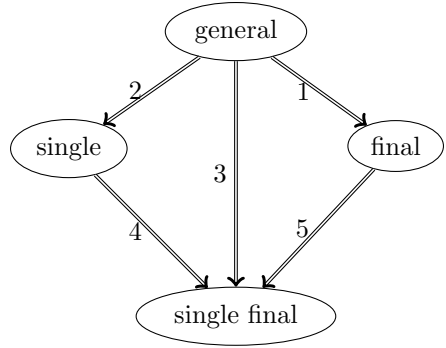

When considering the same norms, we can show the following relations directly using the definitions of the variations of $l_{p,q}$ string stability and the definition of the signal norm, see also \cref{fig:networkedSystems:cent-capt-over} for a visual overview.
\begin{enumerate}

\item General $l_{p,q}$ string stability implies final $l_{p,q}$ string stability.
  \begin{proof}
    General $l_{p,q}$ stability implies that for any bounded input $z$ the output $y$ is bounded, \ac{ie} using the $l_{q}$ signal norm we find that 
    \begin{equation}
      \label{eq:networkedSystems:general-to-final}
      \norm[q]{y} \eqdef \left(\Int[-\infty][\infty]{\sum_{i}\betrag{y_{i}}^{q}}{t}\right)^{1/q}
    \end{equation}
    is bounded.
    Then, \cref{eq:networkedSystems:general-to-final} is an upper bound for
    \begin{equation}
      \label{eq:networkedSystems:general-to-final-2}
      \left(\Int[-\infty][\infty]{\betrag{y_{N}}^{q}}{t}\right)^{1/q} = \norm[q]{y_{N}}
    \end{equation}
    and final $l_{p,q}$ stability is shown.
\end{proof}

\item General $l_{p,q}$ string stability implies single $l_{p,q}$ string stability.
  \begin{proof}
    We select as input signals all $z(t)$ where all besides one entry is zero, \acs{wlg} we say this input is $i$.  Note that for such a signal $\norm[p]{z} = \norm[p]{z_{i}}$. Then, general $l_{p,q}$ string stability guarantees the existence of $R$ independent of $N$ such that the output $\norm[q]{y}$ is bounded for any input signal such that $\norm[p]{z}< \infty$, which shows single $l_{p,q}$ string stability since the input considered is a particular case.
  \end{proof}

\item General $l_{p,q}$ string stability implies single final $l_{p,q}$ string stability.
  \begin{proof}
    We select as input signals all $z(t)$ where all besides one entry is zero, \acs{wlg} we say this input is $i$.  Note that for such a signal $\norm[p]{z} = \norm[p]{z_{i}}$. Then, general $l_{p,q}$ string stability guarantees the existence of $R$ independent of $N$ such that $\norm[q]{y}<R$. Since $\norm[q]{y}$ is an upper bound for $\norm[q]{y_{N}}$, this implies single final $l_{p,q}$ string stability.
  \end{proof}

\item Single $l_{p,q}$ string stability implies single final $l_{p,q}$ string stability.
  \begin{proof}
    According to the definition of single $l_{p,q}$ string stability, any bounded input that acts on a single vehicle produces a bounded output, \acs{ie}
    \begin{equation}
      \label{eq:networkedSystems:general-to-final-5}
      \norm[q]{y} \eqdef \left(\Int[-\infty][\infty]{\sum_{i}\betrag{y_{i}}^{q}}{t}\right)^{1/q} < \infty.
    \end{equation}
    Again this is an upper bound for 
    \begin{equation}
      \label{eq:networkedSystems:general-to-final-6}
      \left(\Int[-\infty][\infty]{\betrag{y_{N}}^{q}}{t}\right)^{1/q} \eqdef \norm[q]{y_{n}},
    \end{equation}
    which is the condition for single final $l_{p,q}$ string stability.
  \end{proof}
\item Final $l_{p,q}$ string stability implies single final $l_{p,q}$ string stability.
  \begin{proof}
    The definition of final $l_{p,q}$ string stability implies that for any input $z(t)$, the output $y_{N}$ remains bounded. We select the input such that the input to all vehicles besides one remain equal to zero. For any of these inputs the output $y_{N}$ remains bounded which shows single final string stability.
  \end{proof}
  
\item Single string stability implies single with input $\alpha$ string stability.
  \begin{proof}
    We select the input signal such that $\norm[p]{z_{\alpha}} < \infty$ and $z_{i}(t) = 0$ for all $i \neq \alpha$. Since the system is single string stable this implies that there exists $R$ such that $\norm[q]{y} < R$, which shows single with input $\alpha$ string stability.
  \end{proof}

  \item Single final $l_{p,q}$ string stability implies single with input $\alpha$ final $l_{p,q}$ string stability.
    \begin{proof}
    We select the input signal such that $\norm[p]{z_{\alpha}} < \infty$ and $z_{i}(t) = 0$ for all $i \neq \alpha$. Since the system is single final string stable there exists $R$ such that $\norm[q]{y_{n}} < R$, which shows single with input $\alpha$ final $l_{p,q}$ string stability.
    \end{proof}
\end{enumerate}

An equally important notion of string stability is internal string stability \cip{swaroop1996stringstability,canudas1999separation,cook2005stability,steffi2008diplomarbeit}. For this notion, instead of considering an input-output relation one considers autonomous systems with initial conditions and the relation to the internal states of the system. As for input-output string stability, it is important to note what norm is used. Hence, we define internal $l_{p,q}$ string stability as follows
\begin{definition}
  A networked system is internal $l_{p,q}$ string stable, if there is a point $x^{\ast}$ such that for any initial conditions $x(0)$ with $\norm[p]{x(0)-x^{\ast}} < \infty$, there exists $R$ for all $N \in \N$ such that
  \begin{equation}
    \norm[q]{x(t)-x^{\ast}} \leq R.
  \end{equation}
\end{definition}

In general internal string stability is a very different notion from input-output string stability.
\begin{remark}
Similar to input-output $\lpnorm[p,q]$ string stability different versions are possible depending on the initial conditions and output signals considered, see for example \cit{steffi2008diplomarbeit}.
\end{remark}

Since internal string stability and general string stability are not equivalent notions, these two can be combined, \acs{ie} the initial state must decay and the output remain bounded for all inputs. Such an approach is for example taken in \cit{steffi2014passivity} in terms of $l_{2,\infty}$ string stability and in \cit{ploeg2014stringstability} in terms of $l_{p}$ string stability. Note that especially the latter proposes a very general definition of (strict) string stability in terms of the $l_{p}$ norm that combines internal $l_{p}$ stability with single $l_{p}$ string stability using as investigated input the reference to the first vehicle. 
Here, we will denote this notion input-to-state string stability.
\begin{definition}
  A networked system is input-to-state $l_{p,q}$ string stable, if for any initial condition $x(0)$ with $\norm[p]{x(0) - x^{\ast}}<\infty$ and input $z(t)$ with $\norm[p]{z} < \infty$, there exists $R$ such that
  \begin{equation}
    \label{eq:networkedSystems:input-to-state-string-stability}
    \norm[q]{y(t)} < R.
  \end{equation}
\end{definition}

\begin{remark}
For a linear networked system input-to-state string stability and internal string stability are equivalent, which is a direct consequence of the linearity.
\end{remark}

\begin{remark}
As for the internal and input-output string stability different versions can be considered depending on the initial conditions, the inputs and outputs considered, \acs{eg} \cit{ploeg2014stringstability}. 
\end{remark}

Finally, the last  main distinction is due to the characteristics of the amplification bound, which includes the rate of increase and an enforced decrease. We collected the main notions in \cref{tab:other-string-stability-forms} focusing on input-output general $l_{p,q}$ string stability (that is, no specific choice for the form or location of the considered input/output signals is taken). Analogous definitions can be obtained for all other notions and versions defined in this paper. To emphasize that we talk about the input-output general $l_{p,q}$ string stability as defined in \cref{tab:string-stability}, we denote this as normal $l_{p,q}$ string stability.

\begin{remark}
  The notion of strict string stability is only defined for directed path graphs, that clearly determine a leader follower structure, such as present in uni-directional communication structures. Hence, when considering strict string stability we assume that a single leader and a well defined order of followers can be identified. In this case a condition in terms of transfer functions can be given such that a vehicle platoon is strict $l_{2}$ string stable in the sense of single final if the transfer function from a vehicle $k$ to its follower $k+1$ is smaller than $1$ for any vehicle $k$ \cip{hermannThesis}, \acs{ie}
  \begin{equation}
      \betrag{\frac{Y_{k+1}(\jm \omega)}{Y_{k}(\jm \omega)}} < 1.
    \end{equation}
\end{remark}

The following relations follow directly from the definition of the various forms.
\begin{enumerate}
\item Normal string stability implies harmonic string stability. \label{item:normal-to-harmonic}
  \begin{proof}
    Note that harmonic string stability in fact allows increasing amplification with increasing string length, but bounds the rate to be sub-exponential. Hence, if a system is string stable, \acs{ie} the amplification is bounded independent of the string length it is also harmonic string stable. More formally, given $R_{0}$ such that $\norm[q]{y} < R_{0}$ for all $N \in \N$ it follows that we can find $R_{1}>1$ such that $\norm[q]{y} < R_{1} \leq R_{1}^{N}$. The existence of $R_{0}$ is given due to normal string stability.
  \end{proof}

\item Strict string stability from $\bar{N}$ implies normal string stability. \label{item:strict-to-norma}
  \begin{proof}
    This follows directly from the definition of strict string stability.
  \end{proof}

\item Strict string stability from $\bar{N}$ implies harmonic string stability.
  \begin{proof}
    This is a direct consequence of \cref{item:normal-to-harmonic,item:strict-to-norma}.
  \end{proof}

\end{enumerate}

The concept of weak string stability \cip{steffi2014passivity,rogge2008vehicleplatoons}, includes a concept of locality, which is not present in the other notions, \acs{ie} the input can be selected small enough to guarantee an arbitrarily small bound on the output. 
\begin{remark}

  In \cit{khatir2004platoon} another form of string string stability is introduced denoted eventual string stability. Their definition is that a vehicle platoon is eventual string stable if there exists a vehicle $\bar{N}$ by which the resultant system is string stable for all following vehicles $i \geq \bar{N}$.
This implies, as for the notion of strict string stability, that the considered vehicle platoon has a defined leader and following ordering.
 We believe that this form of stability is covered indirectly by the definitions given in \cref{tab:string-stability,tab:other-string-stability-forms}. 

\end{remark}

\begin{remark}
  Another important characteristic of a vehicle platoon, however not a safety issue, is coherence or rigidity \cip{bamieh2012coherence}, which measures the notion of how well the string resembles a solid object. This phenomena is not directly related to string stability and looks at a macroscopic view of the system. In fact it is possible for a platoon to be string stable without being coherent. In vehicle platoons this phenomena appears as an ``acordeon'' movement, as described in \cit{bamieh2012coherence}.
\end{remark}

\subsection{Selected Contributions}
\label{sec:results}

It is shown that for linear-time invariant systems under certain conditions string instability can occur for strings \cip{seiler2004vehicleStrings}. While in some cases, \acs{eg} a priori known short string length, eventual instability and string instability are not crucial, in many applications these are important indications for the viability of the vehicle platoon system. There are hence, many results analysing stability and string stability as well as finding methods to design controllers that ensure these. Methods that are suggested to avoid string instability include non-linear control structures \cip{yanakiev1998nonlinear}, non-homogeneous control structures \cip{barooah2009mistuning,lestas2007vehiclePlatoon,khatir2004platoon}, relaxing the formation rigidity \cip{swaroop1996stringstability}, and increasing the communication among the vehicles \cip{seiler2004vehicleStrings, rick2010stringinstability, swaroop1999ahs,cook2007vehiclePlatoon}. 

Due to these reasons it is important to obtain results regarding stability and string stability of the platoon in relation to the information exchange as well as find limitations that exist independent of the controller and the communication involved. We here will summarise the main results related to stability and string stability of the various notions of vehicle platoons both for non-cyclic, in a first instance, and cyclic structures. In \cref{sec:gener-netw-syst} we then discuss briefly how the concepts of eventual instability and string instability can be extended to more general networked systems and available results for these systems. 

\subsubsection{Non-cyclic Interconnections}
\label{sec:non-cyclic-results}

Most studies concentrate on cases with linear dynamics where there is solely relative position information available from the immediate predecessor and possibly the follower. If only the predecessors information is used the control is termed uni-directional \cip{seiler2004vehicleStrings,andres2015thesis,rick2010stringinstability,cook2007vehiclePlatoon}, while if the follower is included in the control we speak of bi-directional control \cip{martinec2016assymetric,herman2015fielder,herman2015circular,lestas2007vehiclePlatoon,andres2015thesis,rick2010stringinstability,cook2007vehiclePlatoon,lestas2007vehiclePlatoon,barooah2005string}. The use of local information alone eliminates the need for communication devices on board, since the information can be found using sensors alone. For example, for bi-directional communication, we find a Laplacian of the form  \cip{barooah2005string}
\begin{equation}
\begin{ownMatr}
    1 & -0.5 &  \\
    -0.5 & 1 & -0.5  \\
    & \ddots & \ddots & \ddots \\
    & & -0.5 & 1 & -0.5 \\
    & & & -0.5 & 0.5
  \end{ownMatr}.
\end{equation}
 Note, that the first and last vehicle are special since they do not possess a leader and follower, respectively. 

\begin{remark}
The term directional here does not relate to the previous use in terms of the associated graph being un-directed or directed. In fact a symmetric bi-directional control leads to un-directed graph Laplacians, while directed graph Laplacians are mostly linked to bi-directional but asymmetric control.
\end{remark}

However, string instability is unavoidable for uni-directional controllers in linear vehicle platoons independent of the controller used, as shown in \cit{seiler2004vehicleStrings}. Hence, without adding additional communication among the agents bi-directional controllers are of high interest.

In \cit{barooah2005string} it is shown for a bi-directional controller that if $L(s)$ contains more than $2$ integrators the system is eventual unstable, \acs{ie} it becomes unstable for large $N$. In the case of two integrators they obtain two additional conditions to maintain closed loop stability independent of $N$. The first condition limits the steady state magnitude of the open loop transfer function $L(s)$ disregarding the integrators, \acs{ie} $\frac{N(0)}{D(0)}>0$. The second condition is developed linking the stability of the networked system to the stability of $N$ transfer functions defined as
\begin{equation}
  G_{i}(s) = \frac{1}{1+ \lambda_{i} L(s)},
\end{equation}
where $\lambda_{i}$ are the eigenvalues of the Laplacian $L$. This second condition holds as well for other communication structures, including cyclic structures.

In regard to string stability it has been shown that bi-directional control can significantly improve the issue \cip{rick2010stringinstability,lestas2007vehiclePlatoon}. In fact, by proper selection of the control gains in $K(s)$ these systems are $l_{2}$ string stable \cip{canudas1999separation}. A similar result is obtained in \cit{cook2007vehiclePlatoon}, where it showed that bi-directional control can avoid the need for a velocity dependent spacing policy, even though the inter vehicle spacing will still be dependent on the length of the vehicle string, under appropriate selection of the controller gains, when the jerk of the vehicles should be bounded for comfort reasons. However, with this approach long transients are caused if the string length grows \cip{rick2010stringinstability,andres2015thesis,cook2007vehiclePlatoon,hermannThesis}.

One method to deal with these long transients is the use of asymmetric control, \acs{ie} the control law uses the information from leading and following vehicles differently \cip{barooah2009mistuning,cook2007vehiclePlatoon,herman2015fielder,martinec2016assymetric,hermannThesis}. In \cit{barooah2009mistuning} the authors show that so called mistuning, \acs{ie} the selection of different controller gains for predecessor and follower, can reduce the sensitivity to disturbances in the sense of $l_{2}$ string stability, but does not necessarily achieve string stability. However, their approach assumes that the vehicles know the desired velocity and make use of this information in their controller. Similarly, this allows the authors in \cit{cook2007vehiclePlatoon} to avoid the introduced limitation on the spacing policy completely in the case where the jerk for the vehicles is limited.

\cit{herman2015fielder} extends the results on asymmetric control strategies where the asymmetry is modeled as a different weighting factor in the Laplacian graph. They conclude that a vehicle platoon with eigenvalues of the Laplacian uniformly bounded away from zero is harmonically string unstable independent of the linear controller utilised without further knowledge, such as desired velocity as is used in \cit{barooah2009mistuning}. 

This lead to a condition of positional symmetry, derived  in \cit{martinec2016assymetric} using the wave approach, that if not satisfied leads to string instability for homogeneous agents. This condition does not hold for other forms of asymmetry, for example in velocity feedback, and asymmetric graph structures can improve the performance of the vehicle platoon in terms of the transient behaviour. Their results are related to previous work of the authors in \cit{herman2015platoons} where a cyclic interconnection of vehicles with three integrators is investigated to find results for the non-cyclic interconnection of such vehicles. As in \cit{martinec2016assymetric} positional symmetry has to hold for asymptotic stability and flock stability, which is a form of harmonic string stability that uses velocity as the performance variable, of the non-cyclic interconnection. 

Besides introduction of asymmetry and bi-directional control, the increase in information exchange is suggested to alleviate string instability. For example, \cit{barooah2009mistuning} assumes the knowledge of a desired velocity in combination with bi-directional asymmetric control, whereas \cit{seiler2004vehicleStrings,andres2015thesis} additionally uses the position or velocity of the leader in the controller. While the first approach does not necessarily require additional communication among the vehicles, the latter means that the communication will increase with the platoon length. This additional communication may cause broadcasting delays that need to be considered, which is done in \cit{andres2015thesis}. There, the broadcast of the velocity of the leader vehicle is investigated in detail. In particular they show that leader velocity broadcast provides string stability if the controller gains are selected properly, without the need to know the number of vehicles in the string nor the individual inter-vehicle spacing of other vehicles. Further, for velocity broadcast string stability is not affected by communication delays that are linearly proportional to the position in the string. 

Alternatively, it is possible to increase the communication range locally rather than including leader position or velocity. This approach is investigated in \cit{cook2007vehiclePlatoon} for unidirectional control. While the use of a wider communication range does not avoid string stability issues, they can be lowered considerably. The same holds for bi-directional vehicle strings as shown in \cit{rick2010stringinstability}, which proves that an increased but limited communication range does not eliminate single final $l_{2}$ string instability, even though the amplification can be reduced. 

\begin{remark}
  There is a subtle difference between the two approaches of increasing the communication range locally but keep it bounded and the broadcast of the leader velocity or position. In the latter the communication range is in fact not bounded, since the total communication length increases with growing string length. This means that for a local communication approach to improve string stability the communication range may need to increase with a growing vehicle string. 
\end{remark}

Another commonly accepted way to achieve string stability is the relaxation of the spacing policy. The works reviewed above all aimed for a constant spacing between the vehicles, which is ideal in terms of efficiency. However, the use of a velocity dependent spacing by introduction of a constant time headway can achieve string stability \cip{rick2010stringinstability,cook2007vehiclePlatoon,chien1992timeheadway,steffi2009timeheadway,hermannThesis}. The drawback of this technique is that it increases the actual inter-vehicle spacing for high speeds and hence lowers the efficiency. 

The time headway can be incorporated into the model by changing the reference input to include no longer a constant reference but a velocity dependent part, \acs{ie} the reference itself would become state dependent. This can be incorporated by a change in $L$ instead. The inclusion of this term becomes straightforward, when using the position of the vehicles as measured output and the inter-vehicle distance as reference. In that case as remarked previously the graph Laplacian is selected as the product of two parts, the one that maps the output to the reference $M$ and the part that maps the reference to the control inputs $\Gamma$. In this set-up the constant time headway can be modeled by adding a diagonal matrix $H$ containing the selected time headway of each vehicle to $M$, such that we use $M+sH$ instead. This means that $L$ is itself now dynamic, which requires more detailed analysis methods.

\cit{rick2010stringinstability} concludes that while a time headway for a uni-directional string does not change the string instability qualitatively a large enough time headway can lead to single $l_{2}$ and $l_{\infty}$ string stability of the vehicle platoon \cip{steffi2008diplomarbeit}. Also, in \cit{cook2007vehiclePlatoon} the constraint of the jerk introduces a lower bound on the time headway. The bound on the time headway can be lowered by the increase of the communication range, which also lowers the absolute inter-vehicle spacing. 

In \cit{yanakiev1998nonlinear} the approach of velocity dependent spacing is extended and nonlinear spacing policies are suggested to achieve string stability while decreasing the actual spacing between vehicles.

While most of the above mentioned results consider the homogeneous case, it is important to discuss the implications of non-homogeneous dynamics \cip{rick2010stringinstability}. In some instances it is the approach taken to achieve string stability by using heterogeneous controllers \cip{khatir2004platoon}. For example, in \cit{khatir2004platoon} controllers with linear increasing gains are chosen to achieve eventual string stability. However, in \cit{rick2010stringinstability} it is shown that heterogeneous control in general does not overcome string instability unless the control bandwidths are allowed to diverge with increasing string lengths.

As we found in this section non-cyclic vehicle platoons suffer from string instability, which is present for uni-directional as well as bi-directional communication structures with limited communication range. In the case of bi-directional communication it is important to distinguish between symmetric and asymmetric approaches. The former can actually achieve $l_{2}$ string stability with proper controller selection, however introduces long transients. The latter while reducing the issue of transients does no longer achieve string stability. Hence, the notion of positional symmetry has been introduced meaning that the control based on position is symmetric, while other control inputs, such as velocity are asymmetric. This seems to achieve a good trade off between the two control strategies. The two main approaches that successfully achieve string stability, are an unbounded increase of the communication range, for example by broadcasting the velocity of the leader, and the introduction of other spacing policies, most commonly a velocity dependent policy by maintaining a constant or variable time headway.

\subsubsection{Cyclic Interconnections}
\label{sec:cycl-interc}

In parallel to work reported on the non-cyclic structures research has also focused on cyclic structures. In some cases these cyclic results are used to obtain the above mentioned results on non-cyclic interconnections \cit{herman2015platoons,herman2015circular}. The reason to start with cyclic interconnections is that the analysis simplifies considerably, since the Laplacian is a circulant matrix. For instance in \cit{herman2015circular} the conjecture is used that an unstable cyclic system means that the system using the associated path graph as communication structure is asymptotically or flock unstable, which is a form of harmonic string stability that uses velocity as the performance variable. This conjecture is then used in \cit{herman2015platoons} to obtain the results for path systems.

While for non-cyclic systems stability is normally not critical, cyclic systems do not exhibit generally a stable behaviour. Hence, the investigation of cyclic systems is focused on eventual instability. Only for eventually stable systems the notion of string stability becomes important.

As for the non-cyclic case the communication structure can both be uni-directional \cip{andres2015thesis} or bi-directional, and can include the use of information from the immediate predecessor and successor \cip{andres2015thesis,herman2015circular} or extended communication ranges. In the uni-directional case it is shown in \cit{andres2015thesis} that eventual instability is unavoidable. This aligns with the results in \cit{seiler2004vehicleStrings}, where string instability of a uni-directional string of vehicles is unavoidable. 

Hence, similar measures as for non-cyclic structures can be used, including bi-directional control, both symmetric and asymmetric, increase of the communication range, use of other spacing policies, as well as the use of heterogeneous controllers and non-linear controllers. In most cases the results found for cyclic interconnections reflect the ones found for non-cyclic ones discussed above. 

For example, the relation between a time headway and the friction present in vehicles is investigated in \cit{rogge2008vehicleplatoons}. They find conditions that guarantees stability and string stability, respectively. These conditions rely on the friction present, the time headway and the controller gain. Their model is a simple point mass model including friction. Using these conditions they are able to conclude in agreement with the other results, see \cit{andres2015thesis}, that a frictionless system becomes unstable for large enough $N$ if the time headway is $0$. These reflect results in \cit{steffi2009timeheadway}.

Similarly, in \cit{herman2015circular} the authors show that a positional symmetry has to hold for stability. Further, they extended the result to show that if the vehicles open loop systems contain more than two integrators instability is unavoidable for growing string lengths independent on the symmetry or asymmetry present. This reflects the result in \cit{herman2015platoons,martinec2016assymetric}.

\section{General Networked Systems}
\label{sec:gener-netw-syst}

Similar stability and performance results are sought for general networked systems, for example in areas of formation control. Especially limitations on the performance are very important for the design of controllers. To this end it is important to generalise the notion of string stability to general networked systems, which is for example done in \cit{li2011performance,yadlapalli2006stability}. 

In regard to stability the identical principle of eventual instability can be used for any network. Ideally, similar string stability definitions can also be utilised. In fact, we can reuse the definitions for general, single, and internal $l_{p,q}$ string stability, as well harmonic string stability, without any change, for example \cit{yadlapalli2006stability} uses both single and general $l_{\infty}$ string stability considering as input either the velocity or disturbances. However, other versions and notions cannot be directly applied. For example the version of final string stability does no longer make sense due to the fact that there is no final vehicle. Hence, in \cref{sec:networkedSystems:networked-stability} we will propose generalisations of the string stability notions given in \cref{sec:string-stability-platoons} for networked systems. Afterwards, we will pinpoint to some selected contributions in that direction in \cref{sec:networkedSystems:select-contr}. 

\subsection{Networked stability}
\label{sec:networkedSystems:networked-stability}

One remarkable difference between a vehicle platoon and a more general networked system is that the communication structure as well as the performance measure are more complex. So in the case of a vehicle platoon if the string is increased by one vehicle, this vehicle appends to the last vehicle, and its performance measure is usually well defined as the distance between some of its predecessors and maybe virtual followers. However, in the case of general networks there are usually multiple locations where the additional agent can be inserted, as well as other performance measures that could be considered. Hence, to be able to generalise the notion of string stability to a form of networked stability, the networked system has to grow in a structured and well defined way. Here, we assume this structure as given such that if we mention in the notion of networked stability for all $N \in \N$, it is inherently clear how the network will expand with increasing $N$.

\begin{remark}
  In the future, it might prove useful to include certain structural expansion within the notion of networked stability. 
\end{remark}

We will base our definitions of networked stability, on our notions of string stability in \cref{sec:string-stability-platoons}. In that context we define an input-output $\lpnorm[p,q]$ networked stability with different variations in \cref{tab:networked-stability}. Note that the variations differ in the considered input and outputs and are related with general, single, and final $\lpnorm[p,q]$ string stability. 

Further, the notion of internal and input-to-state string stability can  be translated directly to general networked systems without any changes. The same holds for the notions of harmonic and weak string stability.

The notion of strict string stability has to be more carefully treated. To this extend we limit the notion of strict networked stability, to networks with a single leader and tree structure. Further, we use the term agent in level $k$ as an agent where there are $k-1$ agents between itself and the leader. So for example an agent $i$ following directly the leader is on level $1$, an agent directly following agent $i$ is on level $2$, and so forth. Then, we can define a form of strict networked stability as follows.
\begin{definition}
  A networked system with a single leader and a well defined tree structure, is strict $\lpnorm[p,q]$ networked stable from level $\bar{N}$, if it is $l_{p,q}$ networked stable and in addition for any input signal $z_{}(t)$ with  $\norm[p]{z_{i}} < \infty$ and agents $j$ and $k$, where $j$ is any agent following $i$ on a level higher or equal than $\bar{N}$ and $k$ is a direct follower of $j$,  
\begin{equation}
  \norm[q]{y_{k}} \leq \norm[q]{y_{j}} 
\end{equation}
for all $j, k$. 
\end{definition}

\subsection{Selected Contributions}
\label{sec:networkedSystems:select-contr}

\cit{li2011performance} uses a notion of $l_{2}$ and $l_{2,\infty}$ networked stability. Their goal is to design a controller such that the total system is asymptotically stable  and the $H_{\infty}$ or $H_{2}$ norm of the transfer function from a disturbance to an output is smaller than a design parameter $\gamma$. For this purpose they define performance regions that are based on a parameter $\sigma$ that takes the place of the eigenvalue of the Laplacian in the system. Then, the region spans the values of this parameter that lead to a stable and well performing system in regard to the design parameter $\gamma$. They also find using this approach a limit for the performance of any controller. Their investigations use an un-directed, un-weighted graph as well as state feedback, but holds for arbitrary graph structures. 

An approach based on $l_{\infty}$ string stability is used  in \cit{yadlapalli2006stability}. Their performance variable is the spacing error of the vehicles to a reference vehicle, as well as the relative velocity to this vehicle. Note that even though they use this as a performance indication they do not assume that all vehicles have access to that information. In that regard they show that the system is unstable if the open loop systems of the individual vehicles (here they consider homogeneous control structure with linear dynamics and un-directed information exchange graphs) has more than two integrators, which generalises results found for vehicle strings \cip{barooah2005string,fax2004platoons}. Additionally, they find that at least one vehicle needs to communicate with a considerable part of the complete formation for stability and string stability in their sense. This links well with the use of the leader position to avoid string instability in vehicle platoons \cip{seiler2004vehicleStrings}.

In \cit{tonetti2010performancelimits,tonetti2011performance} classic control principles combined with Mason's Direct rule are used to obtain performance limitations similar in nature to Bode's integral formula. In fact they show that control merely distributes disturbance rejection at low frequencies between agents. This means by improving one agent's disturbance rejection another agent's performance becomes worse. While the cycle influences the low frequency disturbance the Laplacian spectrum influences the peak in the sensitivity function. Hence, they conclude that for disturbance rejection a two-degree of freedom controller should be used. Their analysis is based on arbitrary networks and includes results both for homogeneous and heterogeneous systems. However, they do not consider weighting of the topology.

\section{Conclusion}
\label{sec:conclusion}

The field of networked systems and in particular vehicle strings has been studied in great detail. In this context especially the notion of string stability becomes important, which means that even though a system can be stable the errors will increase unbounded with increasing numbers of participating agents. This property is mostly studied for vehicle strings, where it showed that without any counter measures, such as increasing communication range, velocity dependent spacing policies, \acs{etc}, string instability occurs. While the case of vehicle strings is very particular, similar effects can occur in more general networked systems, as discussed briefly. Hence, the investigations of why and when these occur become extremely important for the design of controllers. 

In this paper we reviewed the main results attained for vehicle strings. Especially, we formalised different variations of string stability that are commonly investigated and illustrated the importance in the distinction between those with some examples. We hope that this will help in the future to unify the way string stability is presented. Further, we summarised briefly  how these definitions could be generalised for any networked system.

Even though the field of string stability and networked systems lead to a wide range of results there are several open problems. These are for example:
\begin{itemize}
\item Infinite communication length: While it has been shown that increased, but limited communication range does not avoid string instability in non-cyclic vehicle strings \cip{rick2010stringinstability}, the use of the leader velocity or position achieves string stability \cip{seiler2004vehicleStrings}. This latter set-up achieves a form of unbounded communication. It is yet to investigate whether a local unbounded communication structure results equally in string stability.
  
\item Incorporation of weighting: Most research assumes equal weights when treating the information from various sources. However, it has been shown that in vehicle strings the use of asymmetry can improve the response \cip{barooah2009mistuning}. In a similar way asymmetric weights in multi-hop vehicle strings might have a benefit in regard to performance. 
  
\item Limitations in control: As mentioned before, for some configurations string instability is unavoidable independent of the linear controller used. It is important to obtain performance limitations for networked systems to understand what is achievable. These limitations should be independent of the controller that is used.

\item Other performance measures: As we mentioned previously, it is important to distinguish between the different measures that are used, \acs{eg} difference between $l_{\infty}$ and $l_{2}$ string stability. While $\lpnorm[2]$ is the most often discussed string stability notion it is important to investigate and verify the results for other measures, such as $l_{\infty}$, $l_{2,\infty}$ or power signal norm. Further, it is essential to establish relations between the different measures for obtaining an overall understanding. 

\item Communication structure: The communication structure can be of great importance in determining eventual stability or string stability. The conjecture used in \cit{herman2015circular} links eventual stability for vehicle platoons with cyclic communication to string stability in platoons with non-cyclic communication. While in this work the notion is investigated for single final string stability, it is desirable to expand such relations between communication structures in a formal manner and other types of string stability. 

\item Uncertainties: The robustness analysis mostly concentrates on disturbance amplification. Another, important aspect however are uncertainties in the model. At present, the approaches to this topic are fragmented, and primarily rely on either restrictive assumptions or are based on small perturbations.

\item Faults and Security: Due to the reliance on sensors, actuators, and partly communication it is also important to venture in the area of fault or irregularity detection and correction. Initially, these irregularities or faults that can be considered are actually a piece of hardware that is defective or not operating normally. On the other hand it should be extended to include security aspects towards malicious acts, for example individuals that aim for an advantage in the case of vehicle platoons. 

\item Generalisation of the concept of string stability: We gave in \cref{sec:networkedSystems:networked-stability} some examples how the concept of string stability could be generalised for generic networked systems. In doing so we assumed that a structured way is used to expand the network if $N$ grows. It is important to investigate the structural needs to retain networked stability and include cases that include complex cyclic or bi-directional structures, this is the case especially for the concept of strict networked stability.
  
\end{itemize}

\newpage
\bibliographystyle{plain}
\bibliography{networkedSystems}

\begin{thebibliography}{10}

\bibitem{bamieh2012coherence}
B.~Bamieh, M.~R. Jovanovic, P.~Mitra, and S.~Patterson.
\newblock Coherence in large-scale networks: {D}imension-dependent limitations
  of local feedback.
\newblock {\em {IEEE} Transactions on Automatic Control}, 57(9):2235--2249,
  September 2012.

\bibitem{bamieh2002distributedcontrol}
B.~Bamieh, F.~Paganini, and M.~A. Dahleh.
\newblock Distributed control of spatially invariant systems.
\newblock {\em {IEEE} Transactions on Automatic Control}, 47(7):1091--1107,
  2002.

\bibitem{barooah2009mistuning}
P.~Barooah, P.~Mehta, and J.~Hespanha.
\newblock Mistuning-based control design to improve closed-loop stability
  margin of vehicular platoons.
\newblock {\em {IEEE} Transactions on Automatic Control}, 54(9):2100--2113,
  2009.

\bibitem{barooah2005string}
Prabir Barooah and Joa\~o~P. Hespanha.
\newblock Error amplification and disturbance propagation in vehicle strings
  with decentralized linear control.
\newblock In {\em Proceedings of the 44th {IEEE} Conference on Decision and
  Control}, pages 4964--4969, 2005.

\bibitem{cantos2014transients}
Carlos~E. Cantos and J.~J.~P. Veerman.
\newblock Transients in the synchronization of oscillator arrays.
\newblock {\em ArXiv e-prints}, 2014.

\bibitem{chien1992timeheadway}
C.~C. Chien and P.~Ioannou.
\newblock Automatic vehicle-following.
\newblock In {\em American Control Conference, 1992}, pages 1748--1752, 1992.

\bibitem{cook2007vehiclePlatoon}
P.~A. Cook.
\newblock Stable control of vehicle convoys for safety and comfort.
\newblock {\em {IEEE} Transactions on Automatic Control}, 52(3):526--531, March
  2007.

\bibitem{cook2005stability}
P.A. Cook.
\newblock Conditions for string stability.
\newblock {\em Systems and Control Letters}, 54(10):991 -- 998, 2005.

\bibitem{canudas1999separation}
C.~Canudas de~Wit and B.~Brogliato.
\newblock Stability issues for vehicle platooning in automated highway systems.
\newblock In {\em Control Applications, 1999. Proceedings of the 1999 {IEEE}
  International Conference on}, volume~2, pages 1377--1382 vol. 2, 1999.

\bibitem{dunbar2006formation}
William~B. Dunbar and Richard~M. Murray.
\newblock Distributed receding horizon control for multi-vehicle formation
  stabilization.
\newblock {\em Automatica}, 42(4):549 -- 558, 2006.

\bibitem{eyre1998stringStability}
J.~EYRE, D.~YANAKIEV, and I.~KANELLAKOPOULOS.
\newblock A simplified framework for string stability analysis of automated
  vehicles.
\newblock {\em Vehicle System Dynamics}, 30(5):375--405, 1998.

\bibitem{fax2004platoons}
J.~A. Fax and R.~M. Murray.
\newblock Information flow and cooperative control of vehicle formations.
\newblock {\em {IEEE} Transactions on Automatic Control}, 49(9):1465--1476,
  September 2004.

\bibitem{herman2015fielder}
I.~Herman, D.~Martinec, Z.~Hur{\'a}k, and M.~{\' S}ebek.
\newblock Nonzero bound on fiedler eigenvalue causes exponential growth of
  h-infinity norm of vehicular platoon.
\newblock {\em {IEEE} Transactions on Automatic Control}, 60(8):2248--2253,
  2015.

\bibitem{hermannThesis}
Ivo Herman.
\newblock {\em Scaling in vehicle platoons}.
\newblock PhD thesis, Czech Technical University in Prague, May 2016.

\bibitem{herman2015platoons}
Ivo Herman, Dan Martinec, and J.J.P. Veerman.
\newblock Transients of platoons with asymmetric and different laplacians.
\newblock {\em Systems \& Control Letters}, 91:28 -- 35, 2016.

\bibitem{herman2015circular}
Ivo Herman, Dan Martinec, J.J.P. Veerman, and Michael Sebek.
\newblock Stability of a circular system with multiple asymmetric laplacians.
\newblock {\em IFAC-PapersOnLine}, 48(22):162 -- 167, 2015.

\bibitem{hespanha:2007:survey}
Jo{\~a}o~P Hespanha, Payam Naghshtabrizi, and Yonggang Xu.
\newblock A survey of recent results in networked control systems.
\newblock {\em Proceedings of the IEEE}, 95(1):138--162, 2007.

\bibitem{khatir2004platoon}
M.~E. Khatir and E.~J. Davidson.
\newblock Bounded stability and eventual string stability of a large platoon of
  vehicles using non-identical controllers.
\newblock In {\em Decision and Control, 2004. CDC. 43rd {IEEE} Conference on},
  volume~1, pages 1111--1116 Vol.1, 2004.

\bibitem{steffi2009timeheadway}
S.~Klinge and R.~H. Middleton.
\newblock Time headway requirements for string stability of homogeneous linear
  unidirectionally connected systems.
\newblock In {\em Decision and Control, 2009 held jointly with the 2009 28th
  Chinese Control Conference. CDC/{CCC} 2009. Proceedings of the 48th {IEEE}
  Conference on}, pages 1992--1997, 2009.

\bibitem{steffi2008diplomarbeit}
Steffi Klinge.
\newblock Stability issues in distributed systems of vehicle platoons.
\newblock Master's thesis, Otto\textendash von\textendash Guericke\textendash
  Universit{\"a}t Magdeburg, Max Planck Institut, Dynamik komplexer technischer
  Systeme, Sandtorstra{\ss}e 1, 39106 Magdeburg, Deutschland, September 2008.

\bibitem{steffi2012thesis}
Steffi Knorn.
\newblock {\em A Two-Dimensional Systems Stability Analysis of Vehicle
  Platoons}.
\newblock PhD thesis, Hamilton Institute, National University of Ireland
  Maynooth, Co. Kildare, Ireland, October 2012.

\bibitem{steffi2014passivity}
Steffi Knorn, Alejandro Donaire, Juan~C. Ag{\"u}ero, and Richard~H. Middleton.
\newblock Passivity-based control for multi-vehicle systems subject to string
  constraints.
\newblock {\em Automatica}, 50(12):3224 -- 3230, 2014.

\bibitem{lestas2007vehiclePlatoon}
I.~Lestas and G.~Vinnicombe.
\newblock Scalability in heterogeneous vehicle platoons.
\newblock In {\em American Control Conference, 2007. {ACC} 07}, pages
  4678--4683, July 2007.

\bibitem{levine1966vehiclePlatoon}
W.~Levine and M.~Athans.
\newblock On the optimal error regulation of a string of moving vehicles.
\newblock {\em {IEEE} Transactions on Automatic Control}, 11(3):355--361, 1966.

\bibitem{li2011performance}
Zhongkui Li, Zhisheng Duan, and Guanrong Chen.
\newblock On and performance regions of multi-agent systems.
\newblock {\em Automatica}, 47(4):797 -- 803, 2011.

\bibitem{martinec2016assymetric}
D.~Martinec, I.~Herman, and M.~Sebek.
\newblock On the necessity of symmetric positional coupling for string
  stability.
\newblock {\em {IEEE} Transactions on Control of Network Systems}, PP(99):1--1,
  2016.

\bibitem{Melzer1971vehiclePlatoon}
S.M. Melzer and B.C. Kuo.
\newblock Optimal regulation of systems described by a countably infinite
  number of objects.
\newblock {\em Automatica}, 7(3):359 -- 366, 1971.

\bibitem{rick2010stringinstability}
R.~H. Middleton and J.~H. Braslavsky.
\newblock String instability in classes of linear time invariant formation
  control with limited communication range.
\newblock {\em {IEEE} Transactions on Automatic Control}, 55(7):1519--1530,
  2010.

\bibitem{moreau2005consensus}
L.~Moreau.
\newblock Stability of multiagent systems with time-dependent communication
  links.
\newblock {\em Automatic Control, {IEEE} Transactions on}, 50(2):169--182,
  2005.

\bibitem{saber07consensus}
R.~Olfati-Saber, J.A. Fax, and R.M. Murray.
\newblock Consensus and cooperation in networked multi-agent systems.
\newblock {\em Proceedings of the IEEE}, 95(1):215--233, Jan 2007.

\bibitem{oncu212stringstability}
S.~{\"O}nc{\"u}, N.~van~de Wouw, W.~P. M.~H. Heemels, and H.~Nijmeijer.
\newblock String stability of interconnected vehicles under communication
  constraints.
\newblock In {\em 2012 {IEEE} 51st {IEEE} Conference on Decision and Control
  (CDC)}, pages 2459--2464, 2012.

\bibitem{peppard1974pid}
L.~Peppard.
\newblock String stability of relative-motion {PID} vehicle control systems.
\newblock {\em {IEEE} Transactions on Automatic Control}, 19(5):579--581, 1974.

\bibitem{andres2014delays}
Andr{\'e}s~A. Peters, Richard~H. Middleton, and Oliver Mason.
\newblock Leader tracking in homogeneous vehicle platoons with broadcast
  delays.
\newblock {\em Automatica}, 50(1):64 -- 74, 2014.

\bibitem{ploeg2014stringstability}
J.~Ploeg, N.~van~de Wouw, and H.~Nijmeijer.
\newblock Lp string stability of cascaded systems: {A}pplication to vehicle
  platooning.
\newblock {\em {IEEE} Transactions on Control Systems Technology},
  22(2):786--793, 2014.

\bibitem{andres2015thesis}
Andr{\'e}s Alejandro~Peters Rivas.
\newblock {\em Stability and String stability Analysis of Formation Control
  Architectures for Platooning}.
\newblock PhD thesis, Hamilton Institute, Maynooth University, Maynooth,
  October 2015.

\bibitem{rogge2008vehicleplatoons}
J.~A. Rogge and D.~Aeyels.
\newblock Vehicle platoons through ring coupling.
\newblock {\em {IEEE} Transactions on Automatic Control}, 53(6):1370--1377,
  2008.

\bibitem{seiler2004vehicleStrings}
P.~Seiler, A.~Pant, and K.~Hedrick.
\newblock Disturbance propagation in vehicle strings.
\newblock {\em {IEEE} Transactions on Automatic Control}, 49(10):1835--1842,
  2004.

\bibitem{seiler2001losses}
P.~Seiler and R.~Sengupta.
\newblock Analysis of communication losses in vehicle control problems.
\newblock In {\em American Control Conference, 2001. Proceedings of the 2001},
  volume~2, pages 1491--1496 vol.2, 2001.

\bibitem{swaroop1996stringstability}
D.~Swaroop and J.~K. Hedrick.
\newblock String stability of interconnected systems.
\newblock {\em {IEEE} Transactions on Automatic Control}, 41(3):349--357, March
  1996.

\bibitem{swaroop1999ahs}
D.~Swaroop and J.~K. Hedrick.
\newblock Constant spacing strategies for platooning in automated highway
  systems.
\newblock {\em ASME. J. Dyn. Sys., Meas., Control.}, 121:462--470, 1999.

\bibitem{tonetti2010performancelimits}
S.~Tonetti and R.~M. Murray.
\newblock Limits on the network sensitivity function for homogeneous
  multi-agent systems on a graph.
\newblock In {\em Proceedings of the 2010 American Control Conference}, pages
  3217--3222, June 2010.

\bibitem{tonetti2011performance}
Stefania Tonetti and Richard~M. Murray.
\newblock Performance of non-homogeneous multi-agent systems on a graph.
\newblock In {\em Proceedings of the 18th {IFAC} World Congress}, 2011.

\bibitem{yadlapalli2006stability}
S.~K. Yadlapalli, S.~Darbha, and K.~R. Rajagopal.
\newblock Information flow and its relation to stability of the motion of
  vehicles in a rigid formation.
\newblock {\em {IEEE} Transactions on Automatic Control}, 51(8):1315--1319,
  2006.

\bibitem{yanakiev1998nonlinear}
D.~Yanakiev and I.~Kanellakopoulos.
\newblock Nonlinear spacing policies for automated heavy-duty vehicles.
\newblock {\em {IEEE} Transactions on Vehicular Technology}, 47(4):1365--1377,
  1998.

\bibitem{yang2006ncs}
T.C. Yang.
\newblock Networked control system: {A} brief survey.
\newblock {\em {IEE} Proceedings - Control Theory and Applications},
  153(4):403--412, jul 2006.

\bibitem{you:2013:ncs_survey}
Ke-You You and Li-Hua Xie.
\newblock Survey of recent progress in networked control systems.
\newblock {\em Acta Automatica Sinica}, 39(2):101 -- 117, 2013.

\bibitem{zelazo2008formation}
D.~Zelazo, A.~Rahmani, J.~Sandhu, and M.~Mesbahi.
\newblock Decentralized formation control via the edge laplacian.
\newblock In {\em 2008 American Control Conference}, pages 783--788, June 2008.

\bibitem{zelazo2007edgeagreement}
D.~Zelazo, Amirreza Rahmani, and Mehran Mesbahi.
\newblock Agreement via the edge laplacian.
\newblock In {\em Decision and Control, 2007 46th {IEEE} Conference on}, pages
  2309--2314, 2007.

\end{thebibliography}

\end{document}